\documentstyle[aps,prl,preprint,epsfig]{revtex} 

\def \Mee          {\mbox{$m_{ee}$}}
\def \Mgg          {\mbox{$m_{\gamma\gamma}$}}
\def \Meeg         {\mbox{$m_{ee\gamma}$}}
\def \Mfour        {\mbox{$m_{ee\gamma\gamma}$}}
\def \dang         {$|\cos(\theta_{\pi})|$}
\def \dphi         {$|\Delta(\phi)|$}
\def \tmin         {$\theta_{\mathrm min}$}

\def \kltopid      {\mbox{$K_L \rightarrow \pi^0 \pi^0_D$}}

\def \kl           {\mbox{$K_L$}}
\def \ks           {\mbox{$K_S$}}
\def \pid          {\mbox{$\pi^0_D$}}
\def \pipm         {\mbox{$\pi^\pm$}}
\def \klpiee       {\mbox{$K_L \rightarrow \pi^0 e^+ e^-$}}
\def \kspiee       {\mbox{$K_S \rightarrow \pi^0 e^+ e^-$}}

\def \klpigg       {\mbox{$K_L \rightarrow \pi^0 \gamma \gamma$}}
\def \kleeg        {\mbox{$K_L \rightarrow e^+ e^- \gamma$}}
\def \kleegg       {\mbox{$K_L \rightarrow e^+ e^- \gamma \gamma$}}
\def \klzzz        {\mbox{$K_L \rightarrow \pi^0 \pi^0 \pi^0$}}
\def \klpmz        {\mbox{$K_L \rightarrow \pi^+ \pi^- \pi^0$}}
\def \klethree     {\mbox{$K_L \rightarrow \pi^\pm e^\mp \nu$}}

\def \klzz         {\mbox{$K_L \rightarrow \pi^0 \pi^0$}}

\def \pitoeeg      {\mbox{$\pi^0 \rightarrow e^+ e^-\gamma$}}

\def \partder#1#2  {\mbox{$\partial #1\over\partial #2$}}
\def \secder#1#2#3 {\mbox{$\partial^2 #1\over\partial #2 \partial #3$}}
\def \bra#1        {\mbox{$\left\langle #1 \right|$}}
\def \ket#1        {\mbox{$\left| #1\right\rangle$}}
\def \VEV#1        {\mbox{$\left\langle #1\right\rangle$}}
\def \braket#1#2   {\mbox{$\left\langle #1 \left| #2\right\rangle$}}

\def \itaptsq      {\mbox{$P_\perp^2$}}
\def \BR#1         {\mbox{${\mathcal B}$(#1)}}  

\def \omg#1        {\mbox{${\mathcal\cal O}$(#1)}}  

\def \cm           {\mbox{\,cm}}

\def \mevcc        {\mbox{\,MeV/c$^2$}}
\def \gev          {\mbox{\,GeV}}
\def \gevc         {\mbox{\,GeV/c}}

\def \vs           {\mbox{\it vs.}}

\def \etal         {\mbox{\it et al.}}

\def \EPJ          {\mbox{European Phys. Journal}}

\def \NIM          {\mbox{Nucl. Instr. Meth.}}
\def \NP           {\mbox{Nucl. Phys.}}
\def \PL           {\mbox{Phys. Lett.}}
\def \PR           {\mbox{Phys. Rev.}}
\def \PRL          {\mbox{Phys. Rev. Lett.}}

\begin{document}
\draft          
\title{Search for the Decay $K_L \rightarrow \pi^0 e^+ e^-$}
\date{\today}
\maketitle

\begin{center}\normalsize\parindent=0.in
A.~Alavi-Harati$^{12}$,
I.F.~Albuquerque$^{10}$,
T.~Alexopoulos$^{12}$,
M.~Arenton$^{11}$,
K.~Arisaka$^2$,
S.~Averitte$^{10}$,
A.R.~Barker$^5$,
L.~Bellantoni$^{7,\dagger}$,
A.~Bellavance$^9$,
J.~Belz$^{10}$,
R.~Ben-David$^7$,
D.R.~Bergman$^{10}$,
E.~Blucher$^4$, 
G.J.~Bock$^7$,
C.~Bown$^4$, 
S.~Bright$^4$,
E.~Cheu$^1$,
S.~Childress$^7$,
R.~Coleman$^7$,
M.D.~Corcoran$^9$,
G.~Corti$^{11}$, 
B.~Cox$^{11}$,
M.B.~Crisler$^7$,
A.R.~Erwin$^{12}$,
R.~Ford$^7$,
P.M.~Fordyce$^5$,
A.~Glazov$^4$,
A.~Golossanov$^{11}$,
G.~Graham$^4$, 
J.~Graham$^4$,
K.~Hagan$^{11}$,
E.~Halkiadakis$^{10}$,
K.~Hanagaki$^8$,  
S.~Hidaka$^8$,
Y.B.~Hsiung$^7$,
V.~Jejer$^{11}$,
J.~Jennings$^2$,
D.A.~Jensen$^7$,
R.~Kessler$^4$,
H.G.E.~Kobrak$^{3}$,
J.~LaDue$^5$,
A.~Lath$^{10}$,
A.~Ledovskoy$^{11}$,
P.L.~McBride$^7$,
A.P.~McManus$^{11}$,
P.~Mikelsons$^5$,
E.~Monnier$^{4,*}$,
T.~Nakaya$^7$,
K.S.~Nelson$^{11}$,
H.~Nguyen$^7$,
V.~O'Dell$^7$, 
M.~Pang$^7$, 
R.~Pordes$^7$,
V.~Prasad$^4$, 
C.~Qiao$^4$, 
B.~Quinn$^4$,
E.J.~Ramberg$^7$, 
R.E.~Ray$^7$,
A.~Roodman$^4$, 
M.~Sadamoto$^8$, 
S.~Schnetzer$^{10}$,
K.~Senyo$^8$, 
P.~Shanahan$^7$,
P.S.~Shawhan$^4$,
W.~Slater$^2$,
N.~Solomey$^4$,
S.V.~Somalwar$^{10}$, 
R.L.~Stone$^{10}$, 
I.~Suzuki$^8$,
E.C.~Swallow$^{4,6}$,
R.A.~Swanson$^{3}$,
S.A.~Taegar$^1$,
R.J.~Tesarek$^{10}$, 
G.B.~Thomson$^{10}$,
P.A.~Toale$^5$,
A.~Tripathi$^2$,
R.~Tschirhart$^7$, 
Y.W.~Wah$^4$,
J.~Wang$^1$,
H.B.~White$^7$, 
J.~Whitmore$^7$,
B.~Winstein$^4$, 
R.~Winston$^4$, 
T.~Yamanaka$^8$,
E.D.~Zimmerman$^4$
\vspace*{0.1in}

{\bf (The KTeV Collaboration)}\\
\vspace*{0.1in}

$^1$ University of Arizona, Tucson, AZ 85721 \\
$^2$ University of California at Los Angeles, Los Angeles, CA 90095 \\
$^{3}$ University of California at San Diego, La Jolla, CA 92093 \\
$^4$ The Enrico Fermi Institute, The University of Chicago, 
Chicago, IL 60637 \\
$^5$ University of Colorado, Boulder, CO 80309 \\
$^6$ Elmhurst College, Elmhurst, IL 60126 \\
$^7$ Fermi National Accelerator Laboratory, Batavia, IL 60510 \\
$^8$ Osaka University, Toyonaka, Osaka 560 Japan \\
$^9$ Rice University, Houston, TX 77005 \\
$^{10}$ Rutgers University, Piscataway, NJ 08855 \\
$^{11}$ The Department of Physics and Institute of Nuclear and 
Particle Physics, University of Virginia, 
Charlottesville, VA 22901 \\
$^{12}$ University of Wisconsin, Madison, WI 53706 \\
$^{*}$ On leave from C.P.P. Marseille/C.N.R.S., France \\
$^{\dagger}$ To whom correspondence should be addressed.
bellanto@fnal.gov
\end{center}

\begin{abstract}
We report on a search for the decay \klpiee\ carried out by the
KTeV/E799 experiment at Fermilab.  This decay is expected to have a
significant $CP$\, violating contribution and the measurement of its
branching ratio could support the CKM mechanism for
$CP$\, violation or could point to new physics.  Two events were
observed in the 1997 data with an expected background of
$1.06 \pm 0.41$\ events, and we set an upper limit
\BR{\klpiee} $\,<5.1 \times 10^{-10}$\ at the 90\% confidence level.
\end{abstract}

%
%
%
%
The decay \klpiee\ is interesting for the study of 
$CP$\, violation and can be used to search for new physics.  Within the
Standard Model there are three contributions \cite{ref:DonnyO} to this
decay mode. The first is a directly $CP$\, violating contribution from
electroweak penguin and $W^\pm$\ box diagrams; only loops with
$t$\ quarks, which have amplitudes proportional to $\eta$\ in the
Wolfenstein parameterization \cite{ref:LW}, ultimately contribute.
The second contribution is an indirectly $CP$\, violating contribution
\cite{ref:DonnyNu} due to the {\mbox{$K_1$}}\ component of the \kl.
Third, there is a $CP$\, conserving contribution proceeding through
$\pi^0 \gamma^* \gamma^*$\ intermediate states \cite{ref:vdm}, which
can be estimated from measurements \cite{ref:Sydney} of the
decay \klpigg.  The total branching ratio in the Standard Model is
in the range $(3 \sim 10) \times 10^{-12}$.  Branching ratio
predictions are higher in theories containing exotic
particles that contribute to the penguin and box amplitudes.  In
supersymmetric extensions \cite{ref:hope1} to the Standard Model it is
reasonable for \BR{\klpiee} \ to be as high as $2 \times 10^{-11}$,
and possibly as high as $10^{-10}$.  The existing experimental
branching ratio limit \cite{ref:Debbie} is $4.3 \times 10^{-9}$\, at
the 90\% confidence level.  This Letter presents
an improved limit based on data taken by KTeV in 1997.

The major components of the KTeV detector, described in detail in
reference \cite{ref:Kazu,ref:Peter,ref:Greg}, were a magnetic
spectrometer for charged particle tracking, an electromagnetic
calorimeter, and several veto counters to detect particles leaving the
fiducial volume.  The pure CsI electromagnetic calorimeter had an
energy resolution for photons of
$\sigma(E)/E = 0.45\% \oplus 2\%/\sqrt{E}$, where $E$\ is in\.\gev,
and a $\pi^0$\ mass resolution in \klpmz\ of 1.31\mevcc.  The
calorimeter was also used to identify electrons by comparing the energy
deposit to the track momentum as measured by the spectrometer,
rejecting 99.5\% of all charged pions.  Additional
$\pi / e$\, separation was provided with eight transition radiation
detectors (TRDs) located behind the
spectrometer \cite{ref:Greg,ref:TRD}.  Each TRD consisted of a
polypropylene felt radiator followed by a two-plane multiwire
proportional chamber using an 80/20\% Xe/CO$_2$\ mixture.
Pulse height readings from the planes were compared with pulse height
spectra for pions from \klethree\ decays that had been identified with
kinematic and calorimeter requirements.  Each pulse was assigned a
confidence level for the hypothesis it came from a pion.  The
confidence levels for each of the planes where a hit was associated to
a track were combined to yield an overall confidence level for the pion
hypothesis.  For this analysis, cuts giving a pion rejection factor of 
about 35:1 per pion track were used to maximize signal acceptance.

The trigger required hits in the trigger hodoscopes and in the drift
chambers consistent with two coincident charged particles
passing through the detector.  The trigger system counted the number of
isolated clusters of in-time energy in the calorimeter over
$\sim1$\gev\ with a special processor \cite{ref:HCC}; at least four
such clusters were required.  The total energy deposited in the
calorimeter was required to be over 28\gev.  Events with
hadronic showers in the calorimeter were vetoed, as were events with
activity in the photon veto system.  Events that passed the hardware
trigger requirements were reconstructed online and those that passed
loose event topology and CsI electron identification cuts were recorded
on tape.

The signature used to look for \klpiee\ in the detector was two tracks
from oppositely charged particles with a common vertex that deposited
all their energy in the calorimeter, plus two extra calorimeter clusters
which, if interpreted as photons from the charged vertex, had 
$m_{\gamma\gamma} = m_{\pi^0}$.  The decay \kltopid, where
\pid\ denotes a pion with a subsequent \pitoeeg\ (Dalitz) decay, was
used to verify the acceptance calculation and to measure the number of
\kl\ decays in the data sample.  Its signature was similar to the
signal mode's, but with an additional photon satisfying
\Meeg $= m_{\pi^0}$.

A detailed Monte Carlo simulation was used to determine the acceptances
for the signal and normalization modes, allowing for detector geometry,
trigger requirements, reconstruction efficiencies, and analysis
cuts.  It was also used to study background processes.  The
simulation generated kaon decays with the same energy and 
spatial distributions as the data and allowed for particle scattering
and inelastic interactions with material in the beamline.  At the
highest kaon energies included in this analysis, a non-negligible
\ks\ component appeared in decay volume and contributed to the
\kltopid\ sample; this was reproduced by the simulation and was
taken into account in determining the number of \kl\ decays.

Final event reconstruction was done after offline calibration of the
detector.  Recorded events were required to satisfy basic quality
requirements: the electrons were required to not have struck the
calorimeter near the beam holes and the track to cluster matching had
to be unambiguous.  The reconstructed vertex was required to lie within
the neutral beam and well within the vacuum decay region.  The
reconstructed kaon momentum was required the be between 20.3 and
216 \gevc.  Signals used in the trigger were required to lie within
ranges where the simulation modeled the data well.  The tracks were
required to be more than 1\cm\ apart at the first drift chamber and
have an angle over {\mbox{2.25\,mrad}} in the lab frame.  The 
calorimeter cluster associated with each electron had to have an energy
within $\pm5$\% of the electron's momentum as
measured with the spectrometer.  Events with four clusters (including
the two associated to tracks) found by the trigger cluster counter were
used to search for the signal; events with five such clusters were used
to identify \kltopid.

The \kltopid\ sample was selected with requirements on reconstructed
mass and total squared momentum transverse to the \kl\ flight direction
(\itaptsq), and had background of $(0.439 \pm 0.044)$\%.  Using the
acceptance and branching ratio, we determined that there were
$(263.0 \pm1.5_{\mathrm STAT} \pm15.4_{\mathrm SYS} \pm9.1_{\mathrm BR})
\times10^{9}$\, \kl\ decays between 90 and 160\,m from the target with
\kl\ momentum between 20 and 220\gevc.  The systematic uncertainty only
includes effects not common to both signal and normalization mode, and
the third uncertainty is due to uncertainties in
\BR{\klzz} \ and \BR{\pitoeeg} \ .  The
largest systematic uncertainty, $\pm$4.9\%, was from the change in
acceptance when Monte Carlo events were reweighted to match the
occupancy in the beam region of the first drift chamber for the
normalization mode, where tracks are typically close to each other.
Additionally, we assigned a $\pm$3.1\% uncertainty corresponding to
reasonable variations in the selection criteria.

Understanding and suppressing backgrounds is essential for the signal
mode search.  The first background was \klpmz\ where both
\pipm\ showered in the calorimeter; this was removed by requiring the
reconstructed mass of the event, assuming that the tracks were created
by pions, to be over 520\mevcc.

The second background was \klzz\ and \klzzz\ with subsequent \pitoeeg\
decays.  This was reduced by requiring the mass of the two electron
system to be over 140 \mevcc.  There remained some events with two
\pid\ decays where only one $e^+$\ and one $e^-$\ were reconstructed
with a high mass.  These events could also have accidentally coincident
activity.  We rejected events with \Mee\ over 370\mevcc.  To ensure
that we observed all the \kl\ decay products, we required that the
\itaptsq\ be less than 1000\,(MeV/c)$^2$.

The third major background was \klethree\ with a pion that
showered in the calorimeter, accompanied by coincident activity and/or
photons radiated from the electron.  This background limited previous
searches \cite{ref:Debbie}, but was suppressed here with the TRDs.
Tracks with TRD data that fit the pion hypothesis with nominally
4\% confidence level or less were considered electrons; the actual
rejection factor, measured using a sample of \klethree\ decays, was
about 35:1 for each pion track.  This requirement was applied to both
tracks for both normalization mode and signal mode; in the
normalization mode $94.5\pm2.2$\% of events passing all other selection
criteria passed the TRD requirement.

Background from \kspiee, assuming that
$\Gamma(K_S \to \pi^0e^+e^-) \sim \Gamma(K^+ \to \pi^+e^+e^-)$,
is expected to be \omg{$10^{-3}$} \ events.

\begin{figure}
\caption{\Mgg\ \vs\ \Mfour\ for the data after all selection criteria
except the phase space fiducial cuts designed to suppress \kleegg.  The
boxes shown are described in the text; events within the central boxes
have not been plotted.}
\label{fig:nokine}
\begin{center}
\vspace{-0.10in}
\epsfig{file=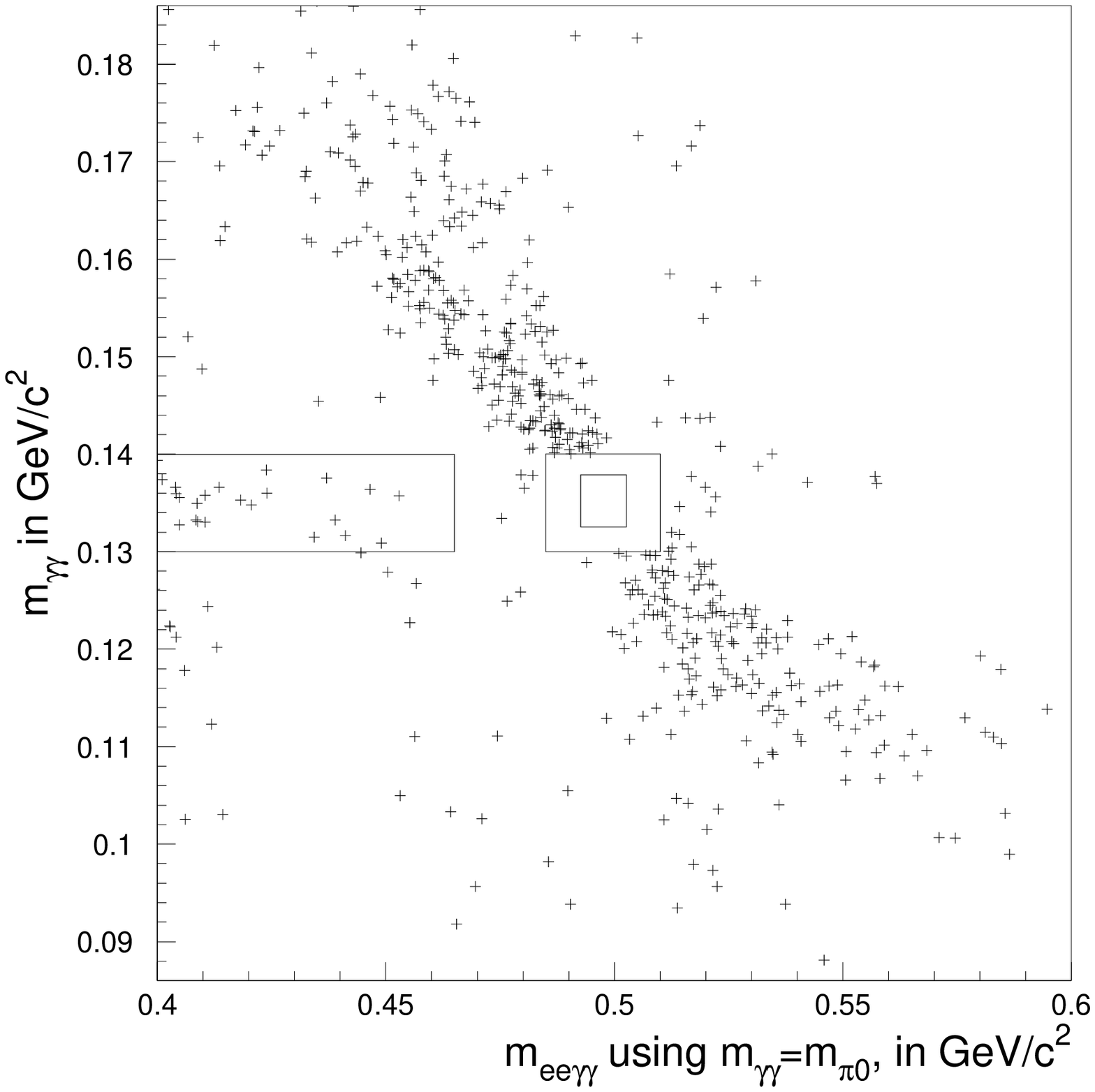,height=5.0in}
\vspace{+0.08in}
\end{center}
\end{figure}

The limiting background was the radiative Dalitz decay \kleegg\ when
\Mgg = $m_{\pi^0}$.  This background was predominantly internal
bremsstrahlung events, but also includes \kleeg\ where an electron
radiated a photon while passing through the detector.  Both components
are treated; details are in reference \cite{ref:Peter}.
Figure \ref{fig:nokine} shows the distribution of
reconstructed two photon mass, \Mgg, \vs\ reconstructed mass of the
four particle system, \Mfour, for data events passing all the selection
criteria described above.  \Mgg\ was calculated assuming that the
photons originated at the charged vertex.  To the contrary,
\Mfour\ was calculated using the $\pi^0$\ mass and measured photon
energies to determine the decay point.  Use of this ``neutral vertex''
improved the \Mfour\ resolution, permitting a tighter search
requirement on this parameter.  The diagonal swath in
Fig. \ref{fig:nokine} is from \kleegg\ decays, where the photons were
incorrectly assumed to have come from a $\pi^0$.

Three regions are marked out in Fig. \ref{fig:nokine}.  To the left
is the region \Mgg$=135\pm5$\mevcc, \Mfour$<465$\mevcc.  Here
the otherwise flat distribution of backgrounds from \klethree\ and
$K_L \rightarrow \pi^0_D \pi^0_D$\ decays with accidental $\pi^0$s
and from $K_L \rightarrow \pi^0 \pi^0_D \pi^0_D$\ decays is peaked due
to the $\pi^0$s in these backgrounds.
This region was not used in estimating the background levels.
The larger central box in Fig. \ref{fig:nokine} is the "blind" region,
(\Mgg$=135\pm5$\mevcc, $485 <$ \Mfour $< 510$\mevcc), which was blanked
out until the analysis procedure and cuts were finalized to avoid
human bias.  This region was also ignored in the background estimation
procedure.  The smaller central box is the signal region
(\Mgg$=135.20\pm2.65$\mevcc, \Mfour = $497.67\pm5.00\mevcc$) the size
of which is roughly a $\pm2\sigma$\ range in the signal Monte
Carlo.

To estimate the background in the signal region, we fit the
distribution of events to
$(B_0+B_\pi m_{\gamma\gamma}+B_K m_{ee\gamma\gamma}+B_1 {\mathcal H})$,
where ${\mathcal H}$\ is the distribution in the (\Mfour,\Mgg) plane of
\kleegg\ as determined from simulation, and $B_i$\ are the parameters
of the fit.  The estimated \kleegg\ background in the signal region of
Fig. \ref{fig:nokine} is $36.9 \pm 2.0$\ events.  The only available
method for suppressing this background is to apply phase space fiducial
cuts \cite{ref:Herb}.  The optimal cuts were found by minimizing the
expected 90\% C.L. limit on \BR{\klpiee} .  That expected branching
ratio limit was calculated using a hypothetical large ensemble of
identical experiments in which the only processes contributing to the
observed number of events were the known backgrounds.  The impact of
the cuts on the signal efficiency, as determined from the simulation,
was taken into account.

The most powerful variables for separating the signal from the
background were \dang, where $\theta_\pi$\ is the angle
between the photons and the kaon in the $\pi^0$\ rest frame, and \tmin,
the minimum angle between any photon and any electron in the kaon rest
frame.  The pion decay angle, which is also the energy
imbalance of the two photons, is a good discriminator because
the comparatively high energy of the photon in a Dalitz decay
causes this angle to be sharply peaked at \dang=1.  In the
signal mode, \dang\ is uniformly distributed because the pion is
spinless.  The minimum electron-photon angle is a good discriminator
because radiated photons usually appear at a small angle to the
electron from which they were emitted.  In the signal, there is
no particular relationship between the photons and the electrons and
\tmin\ is nearly uniformly distributed.  Another widely used
variable is \dphi, the angle between the plane of electrons and the
photons.  Again, this should be peaked near zero for \kleegg\ and
uniformly distributed for \klpiee.  After optimization with \dang\ and
\tmin, no additional rejection was obtained by cutting on \dphi.

The optimal requirements are \dang$<0.788$\ and \tmin$>0.349$; with
these cuts, the expected \kleegg\ background was reduced to
$0.91\pm0.26$\ events with a $\sim24\%$\ signal loss.  The signal
acceptance was $3.609\pm0.017\pm0.085$\%, where the first uncertainty
is from simulation statistics and the second is from the measurement of
TRD efficiency in \kltopid.  The total expected background was
$1.06\pm0.41$\ events, corresponding (in the absence of signal) to an
average expected branching ratio limit of $3.3\times 10^{-10}$.

Several independent checks of the background were made.  One 
important check was to verify that the distribution of the phase space
variables in the diagonal swath of Fig. \ref{fig:nokine} matched
expectations for \kleegg\ decay.  The \kleegg\ branching ratio measured
with this sample was consistent with experimental
results \cite{ref:Peter,ref:TNA48} and theoretical
expectations \cite{ref:Herb}, and the spectra of
\Mee, \Mgg, and other kinematic variables were well simulated.

\begin{figure}
\caption{As in Fig. \ref{fig:nokine}, on an expanded scale around the
signal region, for the data (dots) and signal simulation.  The contours
mark the 1 and 2 $\sigma$\ levels for the signal.  All selection
criteria have been applied.}
\label{fig:box}
\begin{center}
\vspace{-0.10in}
\epsfig{file=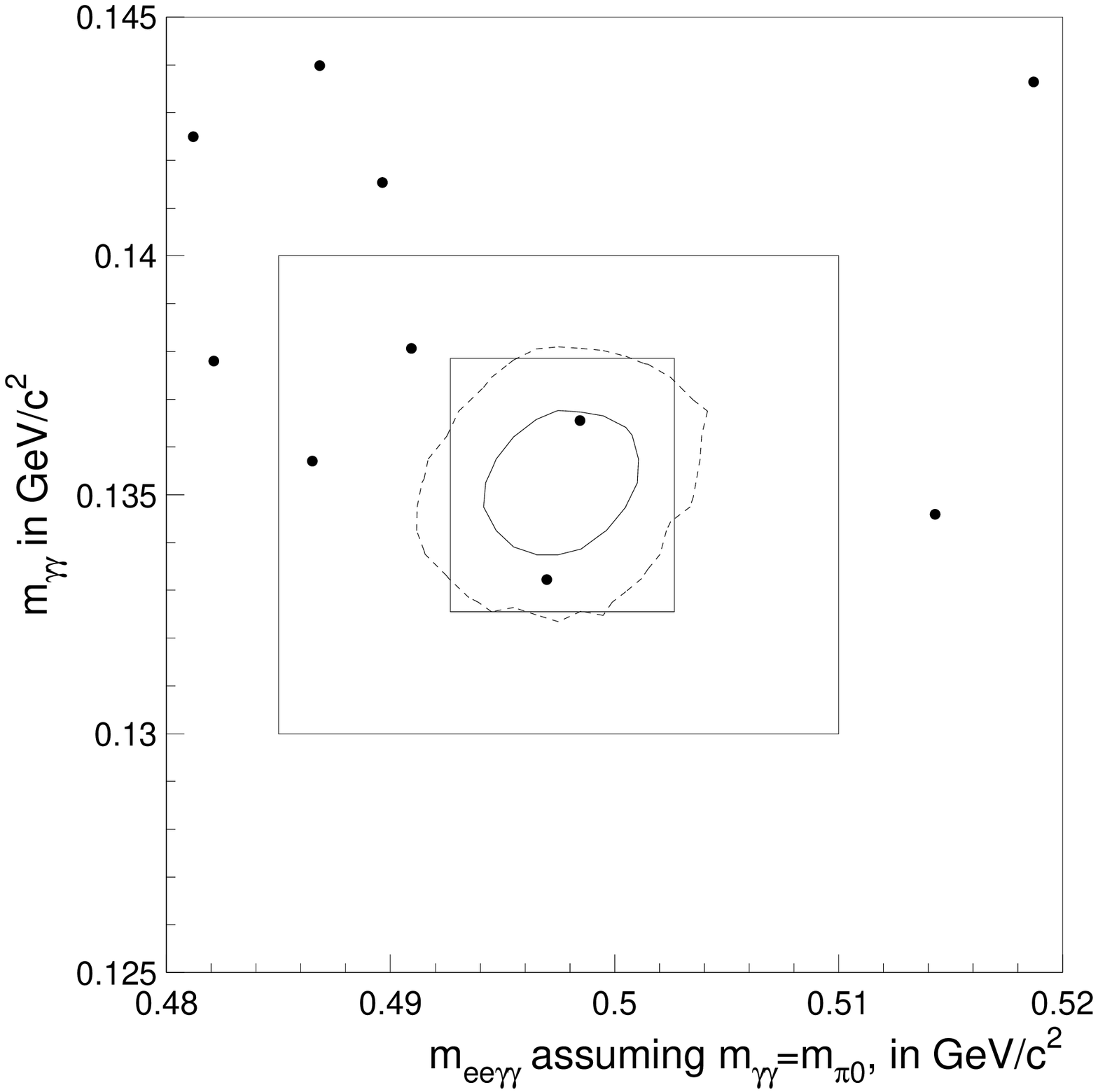,height=5.0in}
\end{center}
\end{figure}

Figure \ref{fig:box} shows \Mgg\ \vs\ \Mfour\ for data, including 
events in the former "blind" region, and the 1 and 2 $\sigma$\ C.L.
contours of the Monte Carlo signal simulation.  Two events exist in the
signal region for the data.  Allowing for the background \cite{ref:FC}
and its uncertainty \cite{ref:Greg}, we set a  90\% C.L. limit of 4.85
events.  With the above acceptance, dataset size, and background level,
we set an upper limit \BR{\klpiee} $\,<5.1 \times 10^{-10}$\ at the
90\% C.L., which is one order of magnitude better than the previous
result.  This limit assumes a uniform three body phase space
distribution for the signal mode.  Alternatively, if the electron pair
is the product of a single vector meson (as in the penguin and indirect
$CP$\, violating terms) and we model the hadronic vertex with a form
factor from \klethree\ decay \cite{ref:Buras,ref:PDG},
$$f_+(q^2) = f_+(0) \{1 + [0.0300 \pm 0.0016](q^2/m_\pi^2) \},$$ 
the 90\% C.L. limit is $\,<5.8 \times 10^{-10}$.

If $r$\ is the fractional contribution to the branching ratio from
direct $CP$\, violation, $|\eta| < 4.4 \sqrt{r}$\ at the 90\% C.L.
While not yet as sensitive in probing the CKM parameterization as B
decays, where recent indirect global analyses \cite{ref:CKMlims} find
$\eta$\ to be below 1, measurements in the kaon system are a valuable
test of whether the same parameterization is valid for both B and K
decays.  With substantially larger event samples, the use of analyses
based on fitting the proper time, Dalitz variables, or decay
asymmetries to observed events \cite{ref:DonnyNu,ref:DebbieT} may
improve this limit.

%
%
We gratefully acknowledge the support and effort of the Fermilab
staff and the technical staffs of the participating institutions for
their vital contributions. This work was supported in part by the U.S.
Department of Energy, The National Science Foundation and The Ministry
of Education and Science of Japan.

\newpage

\end{document}